\documentclass[10pt]{article}
\usepackage{graphicx}
\usepackage{amsmath}
\usepackage{amssymb}
\usepackage{caption2}

\begin{document}
\bibliographystyle{prsty}
\begin{center}
{\large {\bf \sc{  Analysis of  the $X(2370)$ as a glueball based on   rigorous current-field  duality  }}} \\[2mm]
Zhi-Gang Wang\footnote{E-mail: zgwang@aliyun.com.}  \\
 Department of Physics, North China Electric Power University, Baoding 071003, P. R. China
\end{center}

\begin{abstract}
In this work, we take the $X(2370)$ with  $J^{PC}=0^{-+}$ as a glueball consists of three valence gluons, and construct a six-quark current based on  rigorous current-field  duality to obtain the glueball-quark Lagrangian. Then we perform Fierz transformation to  bosonize the  quark current into a series of three pseudoscalar mesons. At last, we obtain ratios among the partial decay widths of the glueball to three pseudoscalar mesons in a model-independent way, which are compatible with the experimental data from the BESIII  Collaboration and support assigning the $X(2370)$ as a glueball.
\end{abstract}



\section{Introduction}
Experimentally, the $J/\psi$ radiative decays are  gluon-rich  and they are believed to be an ideal environment to search for glueballs. We would like to list out the experimental history for the glueball candidate $X(2370)$. In 2011, the BESIII Collaboration  studied  the process  $J/\psi\rightarrow\gamma\pi^{+}\pi^{-}\eta^\prime$ and observed
  the $X(1835)$, $X(2120)$ and  $X(2370)$ in the
$\pi^+\pi^-\eta^\prime$ invariant mass spectrum with statistical significances larger
than $20\sigma$, $7.2\sigma$ and $6.4\sigma$, respectively \cite{BESIII-X2370-PRL-2011}.

In 2019, the BESIII Collaboration  studied the processes
 $J/\psi\to\gamma K\bar{K}\eta^\prime$, and observed the
$X(2370)$ in the $K\bar{K}\eta^\prime$ invariant mass distributions with a statistical significance of 8.3$\sigma$.
The product branching fractions for the decay chains $J/\psi\to \gamma X(2370)\to \gamma K^+K^-\eta^\prime$ and
$\to \gamma X(2370) \to \gamma K^0_S\bar{K}^0_s\eta^\prime$ were  determined to be
$(1.79\pm0.23\pm0.65)\times10^{-5}$ and
$(1.18\pm0.32\pm0.39)\times10^{-5}$, respectively \cite{BESIII-X2370-EPJC-2019}.

In 2021, the BESIII Collaboration studied the process $J/\psi\to\gamma\eta\eta\eta^\prime$ and observed no significant signal for the $X(2370)$   in the $\eta\eta\eta^\prime$ invariant mass distribution \cite{BESIII-X2370-PRD-2021}.

In 2022, the BESIII Collaboration studied  the electromagnetic Dalitz decays $J/\psi \to e^+e^- \pi^+ \pi^- \eta^\prime$, and observed the $X(2120)$ and $X(2370)$ in the $\pi^+ \pi^- \eta^\prime$ invariant mass spectrum with significances of $5.3\sigma$ and $7.3\sigma$, respectively, and measured the corresponding product branching fraction for the $J/\psi\to e^+e^-X(2370)\to e^+e^-\pi^+\pi^-\eta^\prime$ to be $(1.08\pm 0.14 \pm 0.10) \times 10^{-6}$ \cite{BESIII-X2370-PRL-2022}.

Although  the experimental results strongly indicate that the $X(2370)$ is a good candidate of pseudoscalar glueball, it is extremely crucial to   determine its  spin-parity.
In 2024, the BESIII Collaboration performed a partial wave analysis of the decay $J/\psi\rightarrow\gamma K^{0}_{S}K^{0}_{S}\eta^{\prime}$, and observed the $X(2370)$ with the statistical significance  greater than $11.7\sigma$ with  the measured  mass   $2395 \pm 11 {}^{+26}_{-94}\, \mathrm{MeV}$ and width $188^{+18}_{-17} {}^{+124}_{-33}~\mathrm{MeV}$, respectively. In addition, they   determined its quantum numbers to be $J^{PC}=0^{-+}$ for the first time.
The  product branching fraction is ${\rm Br}[J/\psi\rightarrow\gamma X(2370)] \times {\rm Br}[X(2370) \rightarrow f_{0}(980)\eta^{\prime}] \times {\rm Br}[f_{0}(980) \rightarrow K^{0}_{S}K^{0}_{S}] = \left( 1.31 \pm 0.22 ^{+2.85}_{-0.84} \right) \times 10^{-5}$ \cite{BESIII-X2370-PRL-2024}.

In 2026, the BESIII Collaboration studied  the processes  $J/\psi\rightarrow\gamma K^{0}_{S}K^{0}_{S}\pi^{0}$, $\gamma \pi^{0}\pi^{0}\eta$,  and observed  the $X(2370)$  in both the $K^{0}_{S}K^{0}_{S}\pi^{0}$ and $\pi^{0}\pi^{0}\eta$ invariant mass distributions  with statistical significances greater than $14\sigma$ and $20\sigma$, respectively. Combining  measurements from previously reported decay $J/\psi\rightarrow\gamma K^{0}_{S}K^{0}_{S}\eta^{\prime}$,  the BESIII Collaboration determined the mass and width of the $X(2370)$  to be $2359^{+13}_{-14}~\text{MeV}$ and $170^{+44}_{-29}~\text{MeV}$, respectively. Furthermore, the decay chain $X(2370)\to a_{0}^{0}(980)\pi^{0}\to \pi^{0}\pi^{0}\eta$ was observed with a statistical significance exceeding $9\sigma$ \cite{BESIII-X2370-2605}.

Also in 2026, the BESIII Collaboration studied the decay
 $X(2370)\rightarrow K^{*}(892)^{0}\bar{K}^{0}+\mathrm{c.c.}$
 via the process $J/\psi\rightarrow\gamma K_{S}^{0}K_{S}^{0}\pi^{0}$, and observed no evidence, and obtained the product branching fraction  ${\rm Br}[J/\psi\rightarrow\gamma X(2370)] \times {\rm Br}[ X(2370) \rightarrow K^{*}(892)^{0}\bar{K}^{0}+\mathrm{c.c.} \rightarrow K^{0}_{S}K^{0}_{S}\pi^{0}] < 2.7\times 10^{-6}$ at the $90\%$ confidence level \cite{BESIII-X2370-2607}.
The  flavor-singlet meson and  glueball with  $J^{PC}=0^{-+}$ are forbidden to decay into the final state $K^*\bar{K}$  due to the generalized G-parity conservation, which favors the glueball assignment.

 The $X(2370)$ might be  a pseudoscalar glueball consists  of two valence gluons  \cite{ChenY-X2370-G-PRD-2019,CaoJ-X2370-G-PRD-2024,
 LiXT-X2370-G-EPJC-2025,ChenY-X2370-G-2605,ChenHX-X2370-G-PRD-2026}, a conventional pseudoscalar meson, i.e. the fourth radial excitation of the $\eta/\eta^\prime$ \cite{SunZF-X2370-P-PRD-2011,PanCQ-X2370-P-PRD-2020},  a compact hexaquark state \cite{DengCR-X2370-Hexa-PRD-2012}, a compact tetraquark state \cite{ChenHX-X2370-Tetra-EPJC-2020,ChenW-X2370-Tetra-PRD-2025}.
 The spectroscopy of the glueball has been studied extensively by the Lattice QCD \cite{Bali-Spectrum-G-PLB-1993,Morningstar-Spectrum-G-PRD-1999,
  ChenY-Spectrum-G-PRD-2006,Gregory-Spectrum-G-JHEP-2012,
   Mathieu-Spectrum-G-IJMPE-2009,Bali-Spectrum-G-PRD-2000,
   Teper-Spectrum-G-JHEP-2020,Vaccarino-Spectrum-G-PRD-1999} and QCD sum rules \cite{CFQiao-QCDSR-G-PLB-2006,HuangT-QCDSR-G-PRD-1999,
  CFQiao-QCDSR-G-NPB-2016,ChenHX-QCDSR-G-PRD-2021,
  CFQiao-QCDSR-G-PRD-2022,CFQiao-QCDSR-G-PRL-2014,
   Steele-QCDSR-G-NPA-2011,Steele-QCDSR-G-NPA-2003}. The predicted masses vary in large ranges, it is difficult to assign the $X(2370)$ unambiguously based on the mass alone. If the  $X(2370)$ is a pseudoscalar glueball, the decays to three pseudoscalar (P) mesons should be flavor blind, they should be determined by the under-structures, symmetry and kinematical factors.

 In Ref.\cite{DaiLY-X2370-Not-G-PRD-2022}, Sun et al  obtained the   $J/\psi\rightarrow\gamma K^{+} K^{-}\eta'$, $\gamma K_S K_S\eta'$, $\gamma\pi^{+}\pi^{-}\eta'$ and $\gamma\eta\eta\eta'$ decay amplitudes using the chiral effective Lagrangian, and observed
  that the branching ratio ${\rm Br}[J/\psi \to \gamma X(2370)]$ could  be directly extracted from the experiment data ${\rm Br}[J/\psi\to\gamma X(2370) \to \gamma PPP]$. The extracted values ${\rm Br}[J/\psi \to X(2370) \gamma  ]=(2.87\pm0.68)\times 10^{-3}$ or $(3.95\pm0.71)\times 10^{-3}$
  are one order of magnitude  larger than that of  $(0.231\pm0.090)\times 10^{-3}$ or $(0.487\pm 0.143)\times 10^{-3}$ from a Lattice QCD motivated phenomenology analysis.

In Refs.\cite{DaiLY-X2370-Not-G-PRD-2022,
Eshraim-L-PRD-2013,Eshraim-L-PRD-2019,Eshraim-L-EPJC-2023}, the chiral effective  theory is applied to describe the interactions  among  the $X(2370)$ and nonets of the scalar (S)  and pseudoscalar (P) fields,
\begin{align}\label{Chiral-L}
\mathcal{L}=ig_{X}X(\det \Phi -\det\Phi^{\dagger}) \, ,
\end{align}
 with $\Phi=\Phi_0+Z_S S(x) +i Z_P P(x)$, where the $\Phi_0$ is a constant matrix,
  and the $Z_S$ and $Z_P$ are the wave-function renormalization constants.
The chiral Lagrangian in Eq.\eqref{Chiral-L} obeys chiral symmetry and
satisfies Lorentz scalar constraint, but does not have any relation  to
under-structures  of the glueball $X$, irrespective of having two or three valence gluons.

It is interesting to construct an effective Lagrangian by taking account of the under-structures  of the  $X(2370)$ and study the three-body strong decays in a model-independent way.

 The article is arranged as follows:  we obtain partial decay widths of strong decays of the   pseudoscalar glueball in Sect.2;  in Sect.3, we present the numerical results and discussions; and Sect.4 is reserved for our
conclusion.

\section{Strong decays of the   pseudoscalar glueball}

\begin{figure}
\centering
\includegraphics[totalheight=8cm,width=10cm]{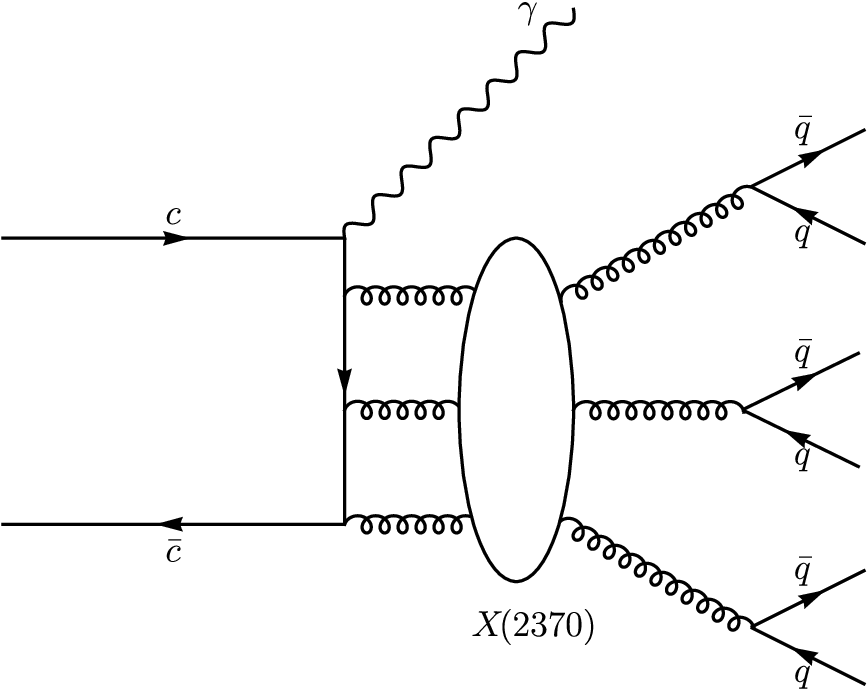}
 \caption{ The mechanism for the decay chain $J/\psi \to \gamma X(2370)\to \gamma PPP$, where the $P$ denotes the pseudoscalar mesons. }\label{Decay-X-fig}
\end{figure}

The $X(2370)$ was observed in the radiative decays  $J/\psi\rightarrow\gamma\pi^{+}\pi^{-}\eta^\prime$,
$\gamma K^+K^-\eta^\prime$,
$ \gamma K^0_S\bar{K}^0_s\eta^\prime$,
$\gamma K^{0}_{S}\bar{K}^{0}_{S}\pi^{0}$, $\gamma \pi^{0}\pi^{0}\eta$
\cite{BESIII-X2370-PRL-2011,BESIII-X2370-EPJC-2019,BESIII-X2370-PRD-2021,
BESIII-X2370-PRL-2022,BESIII-X2370-PRL-2024,BESIII-X2370-2605,BESIII-X2370-2607}, if it is a glueball which consists of three valence gluons, the mechanism for its decays to three
pseudoscalar mesons could be illustrated diagrammatically, see Fig.\ref{Decay-X-fig}.

In the QCD Lagrangian, there exists a term
$g_s\bar{q}_j\frac{\lambda^a_{jk}}{2}\gamma^{\mu}q_k G^a_\mu$, where the $\lambda^a$ are the Gell-Mann matrices in the color space.  The quark current
\begin{eqnarray}\label{Vector-Current}
J_a^\mu(x)&=&\bar{q}_j(x)\frac{\lambda^a_{jk}}{2}\gamma^{\mu}q_k(x) \, ,
 \end{eqnarray}
 and gluon field $ G^a_\mu(x)$ have the same quantum numbers, which lead to a scalar Lagrangian.
 Accordingly, the tensor current
 \begin{eqnarray}
J^a_{\mu\nu}(x)&=&\bar{q}_j(x)
\frac{\lambda^a_{jk}}{2}\sigma_{\mu\nu}q_k(x)\, ,
\end{eqnarray}
and  tensor field $G^a_{\mu\nu}(x)$ have the same quantum numbers,
the dual current
\begin{eqnarray}
\widetilde{J}^a_{\mu\nu}(x)&=&\frac{1}{2}\varepsilon_{\mu\nu\alpha\beta}\bar{q}_j(x)
\frac{\lambda^a_{jk}}{2}\sigma^{\alpha\beta}q_k(x)\, ,
\end{eqnarray}
and  dual tensor field
\begin{eqnarray}
\widetilde{G}^a_{\mu\nu}(x)&=&\frac{1}{2}
\varepsilon_{\mu\nu\alpha\beta}G_a^{\alpha\beta}(x)\, ,
\end{eqnarray}
have the same quantum numbers. Therefore, the pseudoscalar current
 \begin{eqnarray}\label{JJJ-current}
\widetilde{J}(x)&=&f_{abc}\widetilde{J}^a_{\mu\nu}(x)\widetilde{J}_{b}^{\nu\rho}(x)\widetilde{J}_{\rho}^{c\mu}(x)\, ,
\end{eqnarray}
and pseudoscalar field
\begin{eqnarray}\label{GGG-current}
\widetilde{G}(x)&=&f_{abc}\widetilde{G}^a_{\mu\nu}(x)\widetilde{G}_{b}^{\nu\rho}(x)\widetilde{G}_{\rho}^{c\mu}(x)\, ,
\end{eqnarray}
have the same quantum numbers. We construct the effective Lagrangian
\begin{eqnarray}\label{effective-L}
\mathcal{L}(x)&=&\widetilde{g}\widetilde{J}(x)\widetilde{G}(x)\, ,
\end{eqnarray}
 to describe the glueball-quark interactions, which symbolizes  the under-structures of the glueball definitely. Such a method can be applied directly to
 study the glueballs and hybrid states  with a simple replacement
  \cite{CFQiao-QCDSR-G-NPB-2016,ChenHX-QCDSR-G-PRD-2021,
  CFQiao-QCDSR-G-PRD-2022,CFQiao-QCDSR-G-PRL-2014,WangZG-hybrid},
 \begin{eqnarray}
 G^a_{\mu\nu}(x)&\to& J^a_{\mu\nu}(x)\, ,\nonumber\\
 \widetilde{G}^a_{\mu\nu}(x)&\to&\widetilde{J}^a_{\mu\nu}(x)\, ,
 \end{eqnarray}
 to construct the corresponding quark currents therefore the effective Lagrangian.
 In Ref.\cite{ChenHX-X2370-G-PRD-2026}, Tan and  Chen use the current $J^a_\mu(x)$ directly  to construct the effective Lagrangian.

 The field $\widetilde{G}(x)$ couples  potentially to the glueball $\widetilde{G}(p)$ (or $X(2370)$, if the $X(2370)$ is a glueball with three valence gluons),
 \begin{eqnarray}
 \langle 0|\widetilde{G}(0)|\widetilde{G}(p)\rangle&=&\lambda_{\widetilde{G}}\, ,
 \end{eqnarray}
 where the $\lambda_{\widetilde{G}}$ is the  pole residue and can be calculated by the QCD sum rules \cite{CFQiao-QCDSR-G-PLB-2006}.
There exist three additional pseudoscalar fields with $J^{PC}=0^{-+}$ \cite{ChenHX-QCDSR-G-PRD-2021,CFQiao-QCDSR-G-PRD-2022},
\begin{eqnarray}
\widetilde{G}_1(x)&=&f_{abc}\widetilde{G}^a_{\mu\nu}(x)G_{b}^{\nu\rho}(x)G_{\rho}^{c\mu}(x)\, , \nonumber\\
\widetilde{G}_2(x)&=&f_{abc}G^a_{\mu\nu}(x)\widetilde{G}_{b}^{\nu\rho}(x)G_{\rho}^{c\mu}(x)\, , \nonumber\\
\widetilde{G}_3(x)&=&f_{abc}G^a_{\mu\nu}(x)G_{b}^{\nu\rho}(x)\widetilde{G}_{\rho}^{c\mu}(x)\, ,
\end{eqnarray}
which also couple potentially to the pseudoscalar glueball but with the pole residues $\lambda_{\widetilde{G}_k}$, and $k=1$, $2$, $3$. Accordingly, we construct three additional  pseudoscalar currents,
 \begin{eqnarray}
\widetilde{J}_1(x)&=&f_{abc}\widetilde{J}^a_{\mu\nu}(x)J_{b}^{\nu\rho}(x)J_{\rho}^{c\mu}(x)\, ,\nonumber \\
\widetilde{J}_2(x)&=&f_{abc}J^a_{\mu\nu}(x)\widetilde{J}_{b}^{\nu\rho}(x)J_{\rho}^{c\mu}(x)\, ,\nonumber \\
\widetilde{J}_3(x)&=&f_{abc}J^a_{\mu\nu}(x)J_{b}^{\nu\rho}(x)\widetilde{J}_{\rho}^{c\mu}(x)\, .
\end{eqnarray}
Taking  the simple replacements $\widetilde{G}(x) \to \widetilde{G}_k(x)$,
$\widetilde{J}(x) \to \widetilde{J}_k(x)$ and $\widetilde{g} \to \widetilde{g}_k$ with $k=1$, $2$, $3$, we obtain the corresponding Lagrangians.

 We perform Fierz transformation both in the Dirac spinor and color spaces for the current $\widetilde{J}(x)$, the calculation is straightforward but tedious, and we obtain the result,
\begin{eqnarray}\label{Fierz-J}
 -\widetilde{J}&=& \frac{1}{\sqrt{2}}J_{\pi^0}J_{\pi^0}J_{5}^q+\sqrt{2}J_{\pi^+}J_{\pi^-}J_{5}^q
 +\frac{1}{\sqrt{2}}J_{K^+}J_{K^-}J_{5}^q
 +\frac{1}{\sqrt{2}}J_{K^0}J_{\bar{K}^0}J_{5}^q \nonumber\\
 &&+J_{K^+}J_{K^-}J_{5}^s +J_{K^0}J_{\bar{K}^0}J_{5}^s
 +J_{K^+}J_{\bar{K}^0}J_{\pi^-} +J_{K^-}J_{K^0}J_{\pi^+}\nonumber\\
 &&+\frac{1}{\sqrt{2}}J_{K^+}J_{K^-}J_{\pi^0}
 -\frac{1}{\sqrt{2}}J_{K^0}J_{\bar{K}^0}J_{\pi^0}+\frac{1}{3\sqrt{2}}
 J_{5}^qJ_{5}^qJ_{5}^q+\frac{1}{3}
 J_{5}^sJ_{5}^sJ_{5}^s+\cdots\, ,
\end{eqnarray}
where the  $J_{\pi}$ and $J_K$ are the standard pseudoscalar currents interpolating the $\pi$ and $K$ mesons respectively, and
\begin{eqnarray}
J_{5}^q&=&\frac{1}{\sqrt{2}}\left( \bar{u}i\gamma_5u+\bar{d}i\gamma_5d\right)\, ,\nonumber\\
J_{5}^s&=&\bar{s}i\gamma_5s\, .
\end{eqnarray}
We perform Fierz transformation for the currents $\widetilde{J}_{k}(x)$ with $k=1$, $2$, $3$ through  the same procedure, and obtain the result, $\widetilde{J}_k=-\widetilde{J}$.

We usually use the flavor octet and singlet axialvector currents,
\begin{eqnarray}
J_{5\mu}^8&=&\frac{1}{\sqrt{6}}\left(\bar{u}\gamma_\mu\gamma_5u
+\bar{d}\gamma_\mu\gamma_5d -2\bar{s}\gamma_\mu\gamma_5s\right)\, ,\nonumber\\
J_{5\mu}^0&=&\frac{1}{\sqrt{3}}\left(\bar{u}\gamma_\mu\gamma_5u
+\bar{d}\gamma_\mu\gamma_5d +\bar{s}\gamma_\mu\gamma_5s\right)\, ,\
\end{eqnarray}
to study the $\eta-\eta^\prime$ mixing. But it is more convenient to adopt  two independent axialvector currents,
\begin{eqnarray}
J^q_{5\mu}&=&\sqrt{\frac{1}{3}}J_{5\mu}^8+\sqrt{\frac{2}{3}}J_{5\mu}^0=
\frac{1}{\sqrt{2}}\left( \bar{u}\gamma_\mu\gamma_5u+\bar{d}\gamma_\mu\gamma_5d\right)\, , \nonumber\\
J^s_{5\mu}&=&-\sqrt{\frac{2}{3}}J_{5\mu}^8+\sqrt{\frac{1}{3}}J_{5\mu}^0=
\bar{s}\gamma_\mu\gamma_5s\, ,
\end{eqnarray}
which have the couplings,
\begin{eqnarray}
\langle 0|J^q_{5\mu}(0)|\eta(p)\rangle=if_q\cos\phi\, p_{\mu}\, ,\nonumber\\
\langle 0|J^q_{5\mu}(0)|\eta^\prime(p)\rangle=if_q\sin\phi\, p_{\mu}\, ,\nonumber\\
\langle 0|J^s_{5\mu}(0)|\eta(p)\rangle=-if_s\sin\phi\, p_{\mu}\, ,\nonumber\\
\langle 0|J^s_{5\mu}(0)|\eta^\prime(p)\rangle=if_s\cos\phi\, p_{\mu}\, ,
\end{eqnarray}
with the mixing angle $\phi$ \cite{Feldman}.
According to the $U_A(1)$ anomaly,
\begin{eqnarray}
\langle 0|\partial^{\mu}J^q_{5\mu}(0)|\eta(p)\rangle&=& \langle 0|2m_qJ^q_{5}(0)|\eta(p)\rangle
+\sqrt{2}\langle0|\frac{\alpha_s}{4\pi}\widetilde{G}G(0)|\eta(p)\rangle=if_q\cos\phi\, m^2_\eta\, ,\nonumber\\
\langle 0|\partial^{\mu}J^q_{5\mu}(0)|\eta^\prime(p)\rangle&=& \langle 0|2m_qJ^q_{5}(0)|\eta^\prime(p)\rangle
+\sqrt{2}\langle0|\frac{\alpha_s}{4\pi}\widetilde{G}G(0)|\eta^\prime(p)\rangle
=if_q\sin\phi\, m^2_{\eta^\prime}\, ,\nonumber\\
\langle 0|\partial^{\mu}J^s_{5\mu}(0)|\eta(p)\rangle&=& \langle 0|2m_sJ^s_{5}(0)|\eta(p)\rangle
+\langle0|\frac{\alpha_s}{4\pi}\widetilde{G}G(0)|\eta(p)\rangle=-if_s\sin\phi\, m^2_\eta\, ,\nonumber\\
\langle 0|\partial^{\mu}J^s_{5\mu}(0)|\eta^\prime(p)\rangle&=& \langle 0|2m_sJ^s_{5}(0)|\eta^\prime(p)\rangle
+\langle0|\frac{\alpha_s}{4\pi}\widetilde{G}G(0)|\eta^\prime(p)\rangle=if_s\cos\phi\, m^2_{\eta^\prime}\, ,
\end{eqnarray}
we have to subtract the gluon contributions to obtain the hadronic matrix  elements $\langle 0|J_{5}^{q/s}(0)|\eta(p)\rangle$,
$\langle 0|J_{5}^{q/s}(0)|\eta^\prime(p)\rangle$ \cite{Feldman}. And we adopt
the conventional values of the decay constants of the pseudoscalar currents,
\begin{eqnarray}
\langle 0|J_{\pi}(0)|\pi(p)\rangle&=&\frac{f_\pi m_\pi^2}{2m_q}=\mu_{\pi}\, ,\nonumber\\
\langle 0|J_{K}(0)|K(p)\rangle&=&\frac{f_K m_K^2}{m_q+m_s}=\mu_{K}\, ,\nonumber\\
\langle 0|J_{5}^q(0)|\eta(p)\rangle&=&\frac{f_q m_\pi^2\cos\phi}{2m_q}=\mu_{\eta}^q\, ,\nonumber\\
\langle 0|J_{5}^q(0)|\eta^\prime(p)\rangle&=&\frac{f_q m_\pi^2\sin\phi}{2m_q}=\mu_{\eta^\prime}^q\, ,\nonumber\\
\langle 0|J_{5}^s(0)|\eta(p)\rangle&=&-\frac{f_s (2m_K^2-m_\pi^2)\sin\phi}{2m_s}=\mu_{\eta}^s\, ,\nonumber\\
\langle 0|J_{5}^s(0)|\eta^\prime(p)\rangle&=&\frac{f_s (2m_K^2-m_\pi^2)\cos\phi}{2m_s}=\mu_{\eta^\prime}^s\, ,
\end{eqnarray}
with the mixing angle $\phi=39.3^\circ$ \cite{Feldman}.

It is obvious that the current $\widetilde{J}(x)$ couples potentially to three pseudoscalar mesons according to current-meson duality.
Then it is easy to obtain the transition amplitudes $T$ routinely,
\begin{eqnarray}
T_{\widetilde{G}\to \pi^0\pi^0\eta}&=&\sqrt{2}\mu_\pi \mu_\pi \mu_\eta^q \,\lambda_{\widetilde{G}}\,\widetilde{g}\, ,\nonumber\\
T_{\widetilde{G}\to \pi^0\pi^0\eta^\prime}&=&\sqrt{2}\mu_\pi \mu_\pi \mu_{\eta^\prime}^q \,\lambda_{\widetilde{G}}\,\widetilde{g}\, ,\nonumber\\
T_{\widetilde{G}\to \pi^+\pi^-\eta}&=&\sqrt{2}\mu_\pi \mu_\pi \mu_\eta^q \,\lambda_{\widetilde{G}}\,\widetilde{g}\, ,\nonumber\\
T_{\widetilde{G}\to \pi^+\pi^-\eta^\prime}&=&\sqrt{2}\mu_\pi \mu_\pi \mu_{\eta^\prime}^q \,\lambda_{\widetilde{G}}\,\widetilde{g}\, ,
\end{eqnarray}
\begin{eqnarray}
T_{\widetilde{G}\to K^+K^-\eta}&=&\left(\frac{1}{\sqrt{2}}\mu_K \mu_K \mu_{\eta}^q+\mu_K \mu_K \mu_{\eta}^s  \right)\,\lambda_{\widetilde{G}}\,\widetilde{g}\, ,\nonumber\\
T_{\widetilde{G}\to K^+K^-\eta^\prime}&=&\left(\frac{1}{\sqrt{2}}\mu_K \mu_K \mu_{\eta^\prime}^q+\mu_K \mu_K \mu_{\eta^\prime}^s  \right)\,\lambda_{\widetilde{G}}\,\widetilde{g}\, ,\nonumber\\
T_{\widetilde{G}\to K^0\bar{K}^0\eta}&=&\left(\frac{1}{\sqrt{2}}\mu_K \mu_K \mu_{\eta}^q+\mu_K \mu_K \mu_{\eta}^s  \right)\,\lambda_{\widetilde{G}}\,\widetilde{g}\, ,\nonumber\\
T_{\widetilde{G}\to K^0\bar{K}^0\eta^\prime}&=&\left(\frac{1}{\sqrt{2}}\mu_K \mu_K \mu_{\eta^\prime}^q+\mu_K \mu_K \mu_{\eta^\prime}^s  \right)\,\lambda_{\widetilde{G}}\,\widetilde{g}\, ,
\end{eqnarray}
\begin{eqnarray}
T_{\widetilde{G}\to K^+\bar{K}^0\pi^{-}}&=&\mu_K \mu_K \mu_{\pi}\,\lambda_{\widetilde{G}}\,\widetilde{g}\, ,\nonumber\\
T_{\widetilde{G}\to K^-K^0\pi^{+}}&=&\mu_K \mu_K \mu_{\pi}\,\lambda_{\widetilde{G}}\,\widetilde{g}\, ,\nonumber\\
T_{\widetilde{G}\to K^+K^-\pi^{0}}&=&\frac{1}{\sqrt{2}}\mu_K \mu_K \mu_{\pi}\,\lambda_{\widetilde{G}}\,\widetilde{g}\, ,\nonumber\\
T_{\widetilde{G}\to K^0\bar{K}^0\pi^{0}}&=&-\frac{1}{\sqrt{2}}\mu_K \mu_K \mu_{\pi}\,\lambda_{\widetilde{G}}\,\widetilde{g}\, ,
\end{eqnarray}
\begin{eqnarray}
T_{\widetilde{G}\to \eta\eta\eta}&=&\sqrt{2}\mu_\eta^q \mu_\eta^q \mu_{\eta}^q\,\lambda_{\widetilde{G}}\,\widetilde{g}\, ,\nonumber\\
T_{\widetilde{G}\to \eta\eta\eta^\prime}&=&\left(\sqrt{2}\mu_\eta^q \mu_\eta^q \mu_{\eta^\prime}^q+2\mu_\eta^s \mu_\eta^s \mu_{\eta^\prime}^s\right)\,\lambda_{\widetilde{G}}\,\widetilde{g}\, ,
\end{eqnarray}
where the $\widetilde{g}_k$ and $\lambda_{\widetilde{G}_k}$ with $k=1$, $2$, $3$ have been absorbed  in the $\lambda_{\widetilde{G}}\widetilde{g}$.
Then it is straightforward to obtain the partial decay widths according to the formula,
\begin{eqnarray}
\Gamma(\widetilde{G}\to ABC)&=&\frac{1}{\mathcal{S}}\frac{1}{64\pi^3 m_{\widetilde{G}}^2}\int_{(m_A+m_B)^2}^{(m_{\widetilde{G}}-m_C)^2}
  \frac{|T_{\widetilde{G}\to ABC}|^2|\vec{p}_{A}||\vec{p}_{C}|}{s}ds\, ,
\end{eqnarray}
where
\begin{eqnarray}
|\vec{p}_{A}|&=&\frac{\sqrt{\left(s-(m_A+m_B)^2\right)\left(s-(m_A-m_B)^2\right)}}{2\sqrt{s}}\, , \nonumber\\
|\vec{p}_{C}|&=&\frac{\sqrt{\left(m_{\widetilde{G}}^2-(\sqrt{s}+m_C)^2\right)
\left(m_{\widetilde{G}}^2-(\sqrt{s}-m_C)^2\right)}}{2m_{\widetilde{G}}}\, ,
\end{eqnarray}
and the $\mathcal{S}$ are symmetrical  factors of the identical bosons in the final states.

\section{Numerical results and discussions}

We adopt  masses of the pseudoscalar mesons from the Particle Data Group \cite{PDG},
$m_{\pi^0}=134.9768\,\rm{MeV}$,
$m_{\pi^\pm}=139.57039\,\rm{MeV}$,
$m_{K^0}=497.611\,\rm{MeV}$,
$m_{K^\pm}=493.677\,\rm{MeV}$,
$m_{\eta}=547.862\,\rm{MeV}$,
$m_{\eta^\prime}=957.78\,\rm{MeV}$, and take the averages $m_{\pi}=(m_{\pi^0}+m_{\pi^+})/2$, $m_K=(m_{K^+}+m_{K^0})/2$. And we take the decay constants $f_\pi= 130.2\,\rm{MeV}$, $f_K=  155.7\,\rm{MeV}$ from the Particle Data Group \cite{PDG}, $f_q=1.07\, f_\pi$ and $f_s=1.34\, f_\pi$ from the phenomenological analysis
performed on the basis of the Feldmann-Kroll-Stech scheme \cite{Feldman}.

As for the quark masses, we take the isospin limit, $m_q=m_u=m_d$, and adopt  $m_q=-f^2_{\pi}m^2_{\pi}/\left(4\langle \bar{q}q\rangle\right)$ from the Gell-Mann-Oakes-Renner relation and $m_s=27.30\,m_q$ from the Particle Data Group \cite{PDG}. And we take the usually used quark condensate $\langle \bar{q}q\rangle=(240\,\rm{MeV})^3$ \cite{WangZG-Review,ColangeloReview}.

And we set $m_{\widetilde{G}}=m_{X(2370)}=2359\,\rm{MeV}$ \cite{BESIII-X2370-2605} to obtain the numerical values of the partial decay widths,
\begin{eqnarray}
\Gamma( \pi^0\pi^0\eta)&=&5.7264\times 10^{-9}\, {\rm{GeV}}^{13}\lambda_{\widetilde{G}}^2\,\widetilde{g}^2\, ,\nonumber\\
\Gamma( \pi^0\pi^0\eta^\prime)&=&2.0188\times 10^{-9}\, {\rm{GeV}}^{13}\lambda_{\widetilde{G}}^2\,\widetilde{g}^2\, ,\nonumber\\
\Gamma( \pi^+\pi^-\eta)&=&1.1453\times 10^{-8}\, {\rm{GeV}}^{13}\lambda_{\widetilde{G}}^2\,\widetilde{g}^2\, ,\nonumber\\
\Gamma( \pi^+\pi^-\eta^\prime)&=&4.0376\times 10^{-9}\, {\rm{GeV}}^{13}\lambda_{\widetilde{G}}^2\,\widetilde{g}^2\, ,
\end{eqnarray}
\begin{eqnarray}
\Gamma( K^+K^-\eta)&=&1.8051\times 10^{-10}\, {\rm{GeV}}^{13}\lambda_{\widetilde{G}}^2\,\widetilde{g}^2\, ,\nonumber\\
\Gamma( K^+K^-\eta^\prime)&=&2.5830\times 10^{-9}\, {\rm{GeV}}^{13}\lambda_{\widetilde{G}}^2\,\widetilde{g}^2\, ,\nonumber\\
\Gamma( K^0\bar{K}^0\eta)&=&1.8051\times 10^{-10}\, {\rm{GeV}}^{13}\lambda_{\widetilde{G}}^2\,\widetilde{g}^2\, ,\nonumber\\
\Gamma( K^0\bar{K}^0\eta^\prime)&=&2.5830\times 10^{-9}\, {\rm{GeV}}^{13}\lambda_{\widetilde{G}}^2\,\widetilde{g}^2\, ,
\end{eqnarray}
\begin{eqnarray}
\Gamma( K^+\bar{K}^0\pi^{-})&=&8.9358\times 10^{-9}\, {\rm{GeV}}^{13}\lambda_{\widetilde{G}}^2\,\widetilde{g}^2\, ,\nonumber\\
\Gamma( K^-K^0\pi^{+})&=&8.9358\times 10^{-9}\, {\rm{GeV}}^{13}\lambda_{\widetilde{G}}^2\,\widetilde{g}^2\, ,\nonumber\\
\Gamma( K^+K^-\pi^{0})&=&4.4679\times 10^{-9}\, {\rm{GeV}}^{13}\lambda_{\widetilde{G}}^2\,\widetilde{g}^2\, ,\nonumber\\
\Gamma( K^0\bar{K}^0\pi^{0})&=&4.4679\times 10^{-9}\, {\rm{GeV}}^{13}\lambda_{\widetilde{G}}^2\,\widetilde{g}^2\, ,
\end{eqnarray}
\begin{eqnarray}
\Gamma( \eta\eta\eta)&=&2.7427\times 10^{-10}\, {\rm{GeV}}^{13}\lambda_{\widetilde{G}}^2\,\widetilde{g}^2\, ,\nonumber\\
\Gamma( \eta\eta\eta^\prime)&=&8.1253\times 10^{-10}\, {\rm{GeV}}^{13}\lambda_{\widetilde{G}}^2\,\widetilde{g}^2\, .
\end{eqnarray}

Then it is straightforward to obtain the ratios among the partial decay widths,
\begin{eqnarray}\label{BF-GB}
&&\Gamma(\pi\pi\eta):\Gamma(\pi\pi\eta^\prime):
\Gamma( KK\eta):\Gamma(KK\eta^\prime):\Gamma( KK\pi):\Gamma(\eta\eta\eta):\Gamma( \eta\eta\eta^\prime)\nonumber\\
&&=0.641:0.226:0.135:0.193:1.000:0.010:0.030\, ,\nonumber\\
&&=2.08:0.73:0.44:0.63:3.25:0.03:0.10\, .
\end{eqnarray}
Compared with the experimental data from the BESIII  Collaboration \cite{BESIII-X2370-2607}, see Table \ref{x2370summary}, the branching fractions obtained in the present work are very good, which support assigning the $X(2370)$ with the $J^{PC}=0^{-+}$ as a glueball consists of three valence gluons.

\begin{table}[tb]
\centering
\renewcommand\arraystretch{1.2}
\setlength{\tabcolsep}{2.5mm}
\caption{Summary of measured product branching fractions for the $X(2370)$ \cite{BESIII-X2370-2607}.
}
\begin{tabular}{lc}
\hline\hline
Decay channel &${\rm Br}$~$(\times10^{-4})$ \\
\hline

$J/\psi\to\gamma X(2370)\to \gamma K\bar{K}\pi$ & $3.25\pm0.25^{+0.73}_{-0.75}$ \\

$J/\psi\to\gamma X(2370)\to\gamma\pi\pi\eta$ & $3.2\pm0.1^{+0.9}_{-1.0}$ \\

$J/\psi\to\gamma X(2370)\to\gamma\pi\pi\eta^{\prime}$ & $1.94\pm0.04^{+0.33}_{-0.88}$ \\

$J/\psi\to\gamma X(2370)\to \gamma K\bar{K}\eta^{\prime}$ & $0.39\pm0.05\pm0.10$ \\
\hline\hline
\end{tabular}\label{x2370summary}
\end{table}

In Ref.\cite{DaiLY-X2370-Not-G-PRD-2022}, Sun et al  obtained the ratios,
\begin{eqnarray}\label{BF-GB-DaiLY}
&&\Gamma(\pi\pi\eta):\Gamma(\pi\pi\eta^\prime):
\Gamma( KK\eta):\Gamma(KK\eta^\prime):\Gamma( KK\pi):\Gamma(\eta\eta\eta):\Gamma( \eta\eta\eta^\prime)\nonumber\\
&&=2.22:0.16:0.42:0.06:3.25:0.10:0.02\, \, \rm{or} \nonumber\\
&&=2.22:0.17:0.41:0.06:3.25:0.08:0.00\, ,
\end{eqnarray}
from the chiral effective  theory. Compared with  the experimental data from the BESIII  Collaboration \cite{BESIII-X2370-2607}, see Table \ref{x2370summary}, the branching ratios of the channels  $X(2370) \to \pi\pi\eta^\prime$ and $KK\eta^\prime$ are too small as the radiative decay $J/\psi\to \gamma X$ is gluon rich, and disfavor  the glueball assignment. The Lagrangian in Eq.\eqref{Chiral-L} has nothing to do with the under-structures of the pseudoscalar glueball,  irrespective of having two or three valence gluons.

Besides the Fierz transformation shown in Eq.\eqref{Fierz-J}, we can also  transform the current $\widetilde{J}(x)$ into the following form,
\begin{eqnarray}
\widetilde{J}&=&\frac{1}{2\sqrt{2}}J^4_{f_0(500)}J_{qq}+\frac{1}{4}J^4_{a_0(980)}J_{\pi^0}+\frac{1}{4}J^4_{f_0(980)}J_{qq}
+\frac{1}{2\sqrt{2}}J^4_{f_0(980)}J_{ss}+\cdots\, ,
\end{eqnarray}
where
\begin{eqnarray}
J^4_{f_0(500)}&=&\varepsilon^{ijk}\varepsilon^{imn}u^T_jC\gamma_5d_k\, \bar{u}_m\gamma_5C\bar{d}^T_n\, , \nonumber\\
J^4_{a_0(980)}&=&\frac{\varepsilon^{ijk}\varepsilon^{imn}}{\sqrt{2}}\left[u^T_jC\gamma_5s_k\, \bar{u}_m\gamma_5C\bar{s}^T_n -d^T_jC\gamma_5s_k\, \bar{d}_m\gamma_5C\bar{s}^T_n\right]\, , \nonumber\\
J^4_{f_0(980)}&=&\frac{\varepsilon^{ijk}\varepsilon^{imn}}{\sqrt{2}}\left[u^T_jC\gamma_5s_k\, \bar{u}_m\gamma_5C\bar{s}^T_n +d^T_jC\gamma_5s_k\, \bar{d}_m\gamma_5C\bar{s}^T_n\right]\, ,
\end{eqnarray}
the four-quark currents $J^4_{f_0(500)}$, $J^4_{a_0(980)}$ and $J^4_{f_0(980)}$ couple potentially to the tetraquark candidates $f_0(500)$, $a_0(980)$ and $f_0(980)$, respectively. The two-body strong decays,
\begin{eqnarray}
X(2370)&\to& f_0(500)\eta\, , \,\,f_0(500)\eta^{\prime}\, , \,\, a_0(980)\pi^0\, , \,\,f_0(980)\eta\, , \,\,f_0(980)\eta^{\prime}\, , \cdots\, ,
\end{eqnarray}
can take place through  the Okubo-Zweig-Iizuka super-allowed fall-apart mechanism.
In addition, the strong decays,
\begin{eqnarray}
f_0(500)&\to& \pi\pi\, ,\nonumber\\
f_0(980)&\to&\pi\pi\, ,\,\,K\bar{K}\, ,\nonumber\\
a_0(980)&\to&\eta\pi\, ,\,\,K\bar{K}\, , \,\,\eta^\prime\pi\, ,
\end{eqnarray}
are listed in {\it The Review of Particle Physics} \cite{PDG}.
The decay $X(2370)\to a^0_{0}(980)\pi^{0}$ with $a^0_{0}(980)\to \pi^{0}\eta$ was observed with a statistical significance exceeding $9\sigma$ by the BESIII Collaboration \cite{BESIII-X2370-2605}.

However, the QCD sum rules for the $f_0(500)$, $a_0(980)$ and $f_0(980)$ in the tetraquark scenario cannot satisfy the two basic criteria of the QCD sum rules \cite{Scalar-1GeV-Nielsen-PLB-2005-QCDSR,Scalar-1GeV-WangZG-JPG-2005-QCDSR,
Scalar-1GeV-Lee-EPJA-2006-QCDSR}, thus we cannot obtain robust current-tetraquark couplings (or) pole residues to perform quantitative investigations. Furthermore, the light scalar mesons below $1\,\rm{GeV}$ might have both important two-quark and four-quark components, at the present time, we cannot obtain robust predictions for the fractions of these Fock components \cite{Amsler-PRT-2004}.

\section{Conclusion}
In this work, we take the $X(2370)$ with  $J^{PC}=0^{-+}$ as the glueball consists of three valence gluons, and take account of its under-structures to
construct the six-quark current based on  rigorous current-field  duality to obtain the glueball-quark Lagrangian. Then we perform Fierz transformation for the six-quark current both in the Dirac spinor and color spaces to transform  it into a series of color singlet-singlet-singlet type currents, which couple potentially to three pseudoscalar mesons. Then we obtain the transition amplitudes of the glueball to three pseudoscalar mesons, therefore the partial decay widths and ratios among the partial decay widths in a model-independent way, which are compatible with the experimental data from the BESIII   Collaboration and support assigning the $X(2370)$ as the $J^{PC}=0^{-+}$ glueball.

\section*{Acknowledgements}
This  work is supported by National Natural Science Foundation, Grant Number  12575083.

\end{document}